\documentclass[twocolumn,showpacs,preprintnumbers,amsmath,amssymb]{revtex4}

\usepackage{graphicx}
\usepackage{dcolumn}
\usepackage{bm}
\usepackage{color}

\begin{document}

\preprint{APS/123-QED}

\title{Novel universality class for the ferromagnetic transition in the low carrier concentration systems UTeS and USeS exhibiting large negative magnetoresistance\footnote{Phys. Rev. B {\bf 100}, 064413 (2019)}}

\author{Naoyuki Tateiwa$^{1}$}
 \email{tateiwa.naoyuki@jaea.go.jp} 
\author{Yoshinori Haga$^{1}$}%
\author{Hironori Sakai$^{1}$}%
\author{Etsuji Yamamoto$^{1}$}%

\affiliation{
$^{1}$Advanced Science Research Center, Japan Atomic Energy Agency, Tokai, Naka, Ibaraki 319-1195, Japan\\
}
\date{\today}

\begin{abstract}
 We report the novel critical behavior of magnetization in low carrier concentration systems UTeS and USeS that exhibit the large negative magnetoresistance around the ferromagnetic transition temperatures $T_{\rm C}$ $\sim$ 85 and 23 K, respectively. UTeS and USeS crystallize in the same orthorhombic TiNiSi-type crystal structure as those of uranium ferromagnetic superconductors URhGe and UCoGe. We determine the critical exponents, $\beta$ for the spontaneous magnetization $M_{\rm s}$, $\gamma$ for the magnetic susceptibility $\chi$, and $\delta$ for the magnetization isotherm at $T_{\rm C}$ with several methods. The ferromagnetic states in UTeS and USeS have strong uniaxial magnetic anisotropy. However, the critical exponents in the two compounds are different from those in the three-dimensional Ising model with short-range magnetic exchange interactions. Similar sets of the critical exponents have been reported for the uranium ferromagnetic superconductors UGe$_2$ and URhGe, and uranium intermetallic ferromagnets URhSi, UIr and U(Co$_{0.98}$Os$_{0.02}$)Al. The universality class of the ferromagnetic transitions in UTeS and USeS may belong to the same one for the uranium compounds. The novel critical phenomenon associated with the ferromagnetic transition is observed not only in the uranium intermetallic ferromagnets with the itinerant $5f$ electrons but also in the low carrier concentration systems UTeS and USeS with the localized $5f$ electrons. The large negative magnetoresistance in UTeS and USeS, and the superconductivity in UGe$_2$ and URhGe share the similarity of their closeness to the ferromagnetism characterized by the novel critical exponents. 
\end{abstract}

\maketitle
\section{Introduction}
Much interest has been focused on novel physical phenomena in uranium metallic compounds with $5f$ electrons  such as ``hidden order" in URu$_2$Si$_2$, unconventional superconductivity in UPt$_3$ or UBe$_{13}$, and ferromagnetic superconductivity in UGe$_2$, URhGe, and UCoGe\cite{mydosh,pfleiderer1,stewart1,huxley1}. Meanwhile, relatively few studies have been conducted for the magnetism and the electrical conductivity in uranium semimetals or semiconductors. This is in contrast with rare earth magnetic semiconductors, for example, europium chalcogenides EuX (X = O, S, Se, and Te) where the interplay between $4f$ and conduction electrons plays an important role for anomalous physical properties such as a negative magnetoresistance\cite{mauger}.

Very recently, the superconductivity has been discovered in uranium compound UTe$_2$\cite{ran1}. We have studied uranium dichalcogenides UTeS, USeS, and $\beta$-US$_2$ that show the large magnetoresistance at low temperatures\cite{suski,troc,shlyk,ikeda1,ikeda2}. Figure 1 (a) represents the orthorhombic TiNiSi-type crystal structure ($Pnma$) of the uranium dichalcogenides. The structure is the same as those of uranium ferromagnetic superconductors URhGe and UCoGe\cite{huxley1}. Note that UTe$_2$ crystalizes in a different orthorhombic structure ($Immm$)\cite{hutanu}. Figure 1(b) shows the temperature dependencies of the electrical resistivity $\rho$ in the uranium dichalcogenides under magnetic fields of 0 T and 7 T applied parallel to the crystallographic $c$-axis\cite{ikeda1,ikeda2}. The electrical current $J$ was applied along the $b$-axis. UTeS is a semimetal with a low carrier density in the order of $10^{25}$ m$^{-3}$\cite{ikeda1}. USeS and $\beta$-US$_2$ are narrow-gap semiconductors\cite{suski,troc,ikeda2}. UTeS and USeS show ferromagnetic transitions at $T_{\rm C}$ = 85 and 23 K, respectively\cite{troc,ikeda1}. $\beta$-US$_2$ does not order magnetically down to 0.5 K\cite{haga}. Figs. 1(c) and 1(d) show the magnetization at 5.0 K in UTeS and at 2.0 K in USeS, respectively. The ferromagnetic states have strong uniaxial magnetic anisotropy. 

The effect of the magnetic field on the electrical resistivity is strong in the uranium dichalcogenides as shown in Fig. 1(b). In particular, the resistivity values in USeS and $\beta$-US$_2$ decrease largely with increasing field at low temperatures\cite{ikeda2}. The magnitudes of the transverse magnetoresistance are comparable with those in perovskite-type manganese oxides\cite{salamon}. The application of the pressure above 1 GPa induces a ferromagnetic state in $\beta$-US$_2$\cite{tateiwa01}. The large magnetoresistance in the uranium dichalcogenides can be regarded as a novel physical phenomenon that appears around a ferromagnetic phase boundary. The mechanism of the magnetoresistance has not been fully understood yet\cite{ikeda2}. In this paper, we report the novel critical behavior of the magnetization in UTeS and USeS, and its similarity to those in the uranium ferromagnetic superconductors UGe$_2$ and URhGe\cite{tateiwa1}. 

   \begin{figure}[t]
\includegraphics[width=8.3cm]{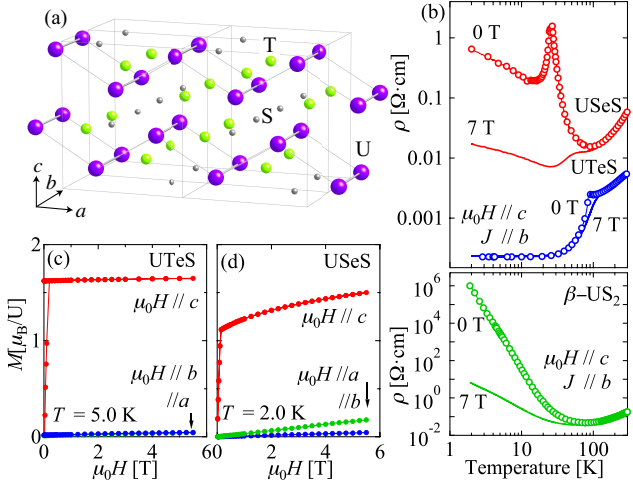}
\caption{\label{fig:epsart}(a)Representation of the orthorhombic TiNiSi-type crystal structure in UTS (T: Te, Se, S). (b)Temperature dependencies of the electrical resistivity $\rho$ under magnetic fields of 0 and 7 T in UTeS\cite{ikeda1}, USeS, and $\beta$-US$_2$\cite{ikeda2}. Magnetic field dependencies of the magnetization in magnetic fields applied along the $a$, $b$, and $c$-axes (c) at 5.0 K in UTeS\cite{ikeda1} and (d) at 2.0 K in USeS.}
\end{figure} 
   \begin{figure}[]
\includegraphics[width=8.5cm]{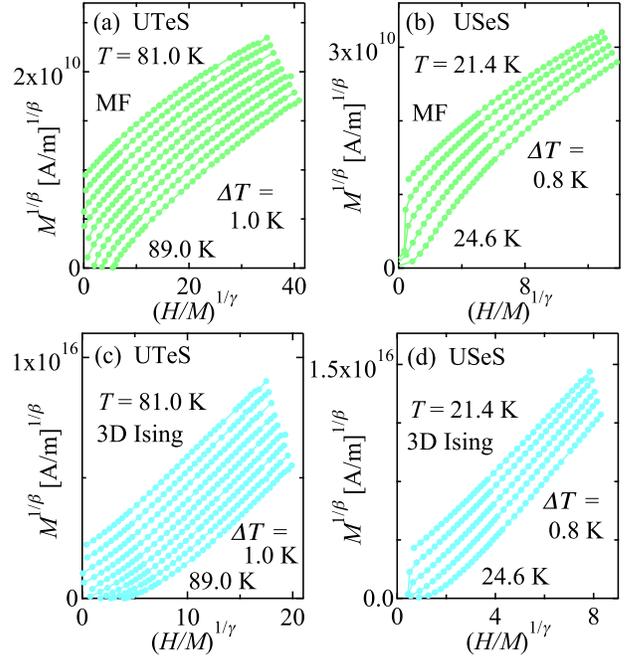}
\caption{\label{fig:epsart} Magnetic isotherms in the form of $M^{1/{\beta}}$ vs. $(H/M)^{1/{\gamma}}$ with the mean field theory ($\beta$ = 0.5 and $\gamma$ = 1.0) (a) for 81.0 K $\le$ $T$ $\le$ 89.0 K in UTeS and (b) for 21.4 K $\le$ $T$ $\le$ 24.6 K in USeS. Isotherms with the short-range (SR) 3D Ising model ($\beta$ = 0.325 and $\gamma$ = 1.241) in (c) UTeS and in (d) USeS.}
\end{figure} 
\section{Experiment}
Single crystals of UTeS and USeS were grown by the chemical transport using bromine as a transport agent\cite{ikeda1,ikeda2}. Neither impurity phase nor off-stoichiometric chemical composition larger than about 1$\%$ was detected in the X-ray diffraction and the Electron Probe Micro Analysis (EPMA). Magnetization was measured in a commercial superconducting quantum interference device (SQUID) magnetometer (MPMS, Quantum Design). The magnetic field ${{\mu}_0}{H_{\rm ext}}$ was applied along the magnetic easy $c$ axis in the orthorhombic structure. We determine the internal magnetic field ${{\mu}_0}H$ by subtracting the demagnetization field $DM$ from ${{\mu}_0}{H_{\rm ext}}$ : ${{\mu}_0}H$ = ${{\mu}_0}{H_{\rm ext}}$ - $DM$. The demagnetization factors $D$ = 0.50 and 0.46 were estimated from the macroscopic dimensions of the single crystals of UTeS and USeS, respectively. 

\section{Results}
In the asymptotic critical region near $T_{\rm C}$ where the mean field theory fails, the magnetic correlation length $\xi$ = ${\xi}_0$ $|({T}-{T_{\rm C}})/{T_{\rm C}}|^{-{\nu}}$ diverges. Here, $\nu$ is the critical exponent. The spontaneous magnetization $M_{\rm s}$, the initial susceptibility $\chi$, and the magnetization at $T_{\rm C}$ follow universal scaling laws\cite{privman}.

     \begin{figure}[t]
\includegraphics[width=8.5cm]{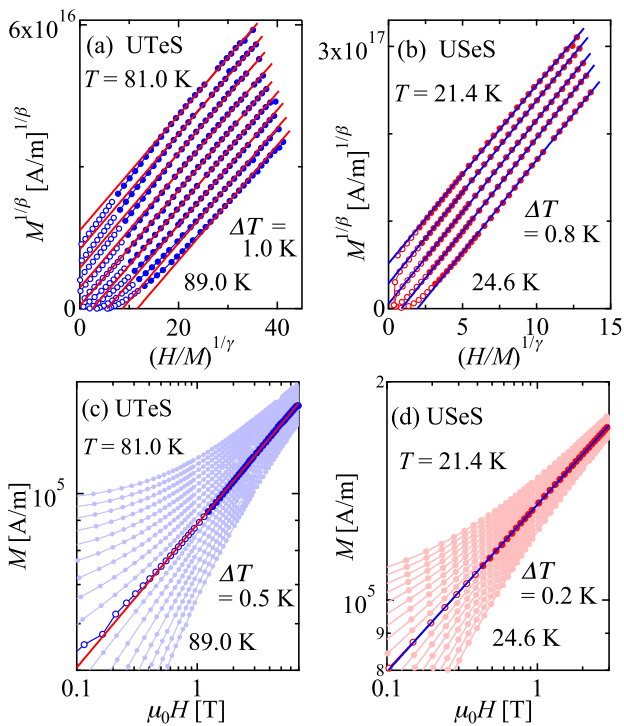}
\caption{\label{fig:epsart}Modified Arrott plot (MAP) of magnetic isotherms (a) for 81.0 K $\le$ $T$ $\le$ 89.0 K in UTeS and (b) for 21.4 K $\le$ $T$ $\le$ 24.6 K in USeS. Data represented as close circles in (a) and (b) are analyzed with the Arrott-Noakes equation of state [Eq. (4)]. Solid lines show fits to the data in the higher magnetic field region with a linear function. Magnetic field dependencies of the magnetization in (c) UTeS and in (d) USeS. Bold circles indicate the critical isotherms at 85.0 K and 23.2 K for UTeS and USeS, respectively. Solid lines represent fits to the data represented as closed bold circles with Eq. (3).}
\end{figure} 

     \begin{figure}[t]
\includegraphics[width=8.5cm]{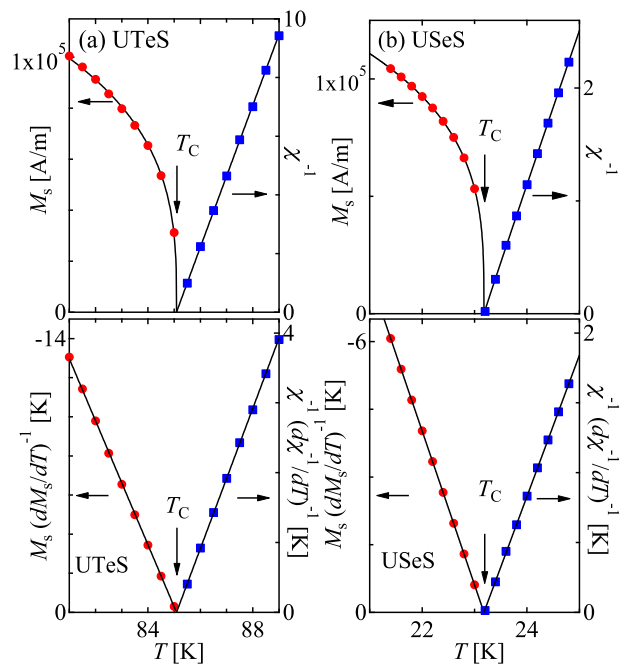}
\caption{\label{fig:epsart}Temperature dependencies of $M{_{\rm s}}(T)$ and ${\chi}^{-1}$ determined from the MAPs (upper panels) and Kouvel-Fisher plots (lower panels) in (a) UTeS and in (b) USeS.}
\end{figure}

    \begin{eqnarray}
 M{_{\rm s}}(T) &{\propto}& |t|{^{\beta}}  {\;}  {\;} {\;}{\;}{\;} (T < T{_{\rm C}})\\ 
{{\chi}}(T){^{-1}}&{\propto}&  {|t|}^{{{\gamma}}'}  {\;}  {\;} (T<T{_{\rm C}}), {\;} {|t|}^{{{\gamma}}} {\;}{\;}(T{_{\rm C}}< T)\\ 
 {M}({{\mu}_0}H)& {\propto} & ({{\mu}_0}H)^{1/{\delta}} {\;} {\;}(T = T{_{\rm C}})
   \end{eqnarray}        
 Here, $t$ is the reduced temperature $t$ = $(T-{T_{\rm C}})/{T_{\rm C}}$. $\beta$, $\gamma$, ${\gamma}'$ and $\delta$ are the critical exponents. 

 Figures 2(a) and 2(b) show the magnetic isotherms in the form of $M^{1/{\beta}}$ versus $(H/M)^{1/{\gamma}}$ with the mean field (MF) theory ($\beta$ = 0.5 and $\gamma$ = 1.0) for 81.0 K $\le$ $T$ $\le$ 89.0 K in UTeS and for 21.4 K $\le$ $T$ $\le$ 24.6 K in USeS, respectively. The data do not form straight lines. This suggests that the mean field theory is not suitable to describe the magnetization around $T_{\rm C}$. Figs. 2(c) and 2(d) represent the isotherms in the form of $M^{1/{\beta}}$ vs. $(H/M)^{1/{\gamma}}$ with the 3D Ising model with short-range (SR) exchange interactions ($\beta$ = 0.325 and $\gamma$ = 1.241) for UTeS and USeS, respectively. The isotherms are curved, suggesting that the 3D Ising model is also not appropriate. 
 
 We analyzed the data with the following Arrott-Noakes equation of state\cite{arrott}:
   \begin{eqnarray}
  &&(H/M){^{1/{\gamma}}} = (T-{T_{\rm C}})/{T_1} + (M/{M_{1}})^{1/{\beta}}
   \end{eqnarray}
, where $T_1$ and $M_1$ are material constants.

 Figures 3(a) and 3(b) show the modified Arrott plot (MAP) in the form of $M^{1/{\beta}}$ vs. $(H/M)^{1/{\gamma}}$ for UTeS and USeS, respectively. The isotherms become straight if the appropriate values of $T_{\rm C}$, $\beta$, and $\gamma$ are chosen. We determine these parameters from fits of Eq. (4) to the data for 81.0 K $\le$ $T$ $\le$ 89.0 K and 1.2 T $\le$ ${{\mu}_0}H$ $\le$ 7.0 T in UTeS, and those for 21.4 K $\le$ $T$ $\le$ 24.6 K and 0.4 T $\le$ ${{\mu}_0}H$ $\le$ 3.0 T in USeS. The values of $T_{\rm C}$, $\beta$, and $\gamma$ are determined as $T_{\rm C}$ = 84.88 $\pm$ 0.05 K, $\beta$ = 0.309 $\pm$ 0.003, and ${\gamma}$ = 0.998 $\pm$ 0.003 for UTeS, and $T_{\rm C}$ = 23.09 $\pm$ 0.03 K, $\beta$ = 0.300 $\pm$ 0.003, and ${\gamma}$ = 1.00 $\pm$ 0.02 for USeS. The parameters are shown in Table I. The data used for the analyses are represented as closed circles in Figs. 3(a) and 3(b). The data points in the MAPs generally form straight lines but those in the low magnetic field region deviate from the lines. There are several reasons for the deviation such as the movement of domain walls or sample inhomogeneities. In addition, there might be an error in the calculated value of the demagnetization factor $D$. The origins of the deviation have been discussed in Ref. 19, although it has not been completely understood yet. Solid lines in Figs. 3(a) and 3(b) represent fits to the data in the high magnetic field region with a linear function in order to obtain the spontaneous magnetic moment $M_{\rm s}$  and the magnetic susceptibility $\chi$. The temperature dependencies of the quantities will be used in the analysis with the Kouvel-Fisher method.

We determine the critical exponent $\delta$ from the critical isotherm at $T_{\rm C}$. The value of $\delta$ is determined as $\delta$ = 4.21 $\pm$ 0.04 for UTeS and 4.34 $\pm$ 0.04 for USeS from fits to the isotherms at 85.0 K for UTeS and at 23.2 K for USeS with Eq. (3) as shown in Figs. 4(c) and 4(d), respectively. The data shown as closed circles are analyzed. The exponents $\beta$, $\gamma$, and $\delta$ should be related by the Widom scaling law $\delta$ = 1+$\gamma$/$\beta$\cite{widom}. The value of $\delta$ is estimated as 4.23 $\pm$ 0.06 for UTeS and 4.33 $\pm$ 0.10 for USeS using the $\beta$ and $\gamma$ values in the MAPs. These values are consistent with those determined from the critical isotherms. 
 
 Next, we analyze the data with the Kouvel-Fisher method where the critical exponents can be determined more accurately\cite{kouvel}. The solid lines in Figs 3(a) and 3(b) intersect with the vertical axis at $M^{1/{\beta}}$ = $M_{\rm s}^{1/{\beta}}$ for $T$ $<$ $T_{\rm C}$ and with the transverse axis at $(H/M)^{1/{\gamma}}$ = $(1/{\chi})^{1/{\gamma}}$ for $T_{\rm C}$ $<$ $T$. The values of $M_{\rm s}(T)$ and ${\chi}(T)$ can be obtained by inserting the $\beta$ and $\gamma$ values. Figures 4(a) and 4(b) show the temperature dependencies of $M_{\rm s}(T)$ and ${\chi}^{-1}(T)$ for UTeS and USeS, respectively. Solid lines represent fits to the data with Eqs. (1) and (2). We determine the critical exponents $\beta$ and $\gamma$ with the Kouvel-Fisher (KF) method where temperature-dependent exponents ${\beta}(T)$ and ${\gamma}(T)$ are introduced as follows\cite{kouvel}:

      \begin{figure}[t]
\includegraphics[width=8.5 cm]{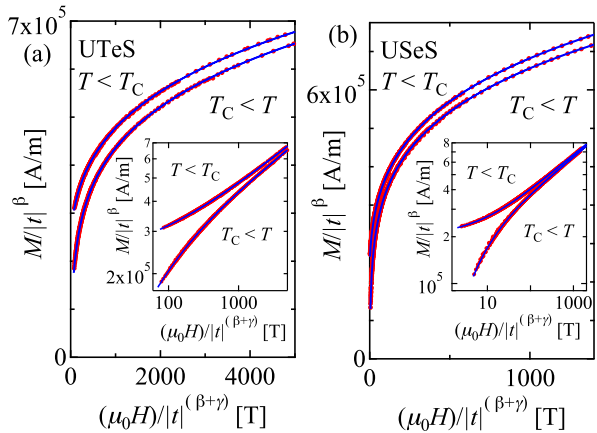}
\caption{\label{fig:epsart}Renormalized magnetization $m$ as a function of renormalized field $h$ following Eq. (7) below and above $T_{\rm C}$ for (a) UTeS and for (b) USeS. Solid lines represent best-fit polynomials. }
\end{figure} 

 \begin{eqnarray}
 M{_{\rm s}}(T)[dM{_{\rm s}}(T)/dT]{^{-1}} &=& (T-T{_{\rm C}}^{-})/{\beta}(T)\\ 
  {{\chi}}{^{-1}}(T)[d {{\chi}}{^{-1}}(T)/dT]{^{-1}} &=& (T-T{_{\rm C}}^{+})/{\gamma}(T)
  \end{eqnarray}

  The quantities ${\beta}(T)$ and ${\gamma}(T)$ are equal to the critical exponents $\beta$ and $\gamma$, respectively, in the limits $T$ $\rightarrow$ $T_{\rm C}$ and $H$ $\rightarrow$ 0. The slopes of $M{_{\rm s}}(T)[dM{_{\rm s}}(T)/dT]{^{-1}}$ and $ {{\chi}}{^{-1}}(T)[d {{\chi}}{^{-1}}(T)/dT]{^{-1}}$-$T$ plots at $T_{\rm C}$ yield the $\beta$ and $\gamma$ values, respectively. The fits to the data with Eqs. (5) and (6) are shown as solid lines in the low panels of Figs. 4(e) and 4(f), respectively. The parameters are determined as $\beta$ = 0.315 $\pm$ 0.003 and $\gamma$ = 0.996 $\pm$ 0.003, and $T_{\rm C}$ = 85.09 $\pm$ 0.04 K for UTeS, and  $\beta$ = 0.293 $\pm$ 0.003, $\gamma$ = 0.989 $\pm$ 0.003, and $T_{\rm C}$ = 23.18 $\pm$ 0.03 K for USeS. Here, $T_{\rm C}$ is defined as $T_{\rm C}$ = ($T{_{\rm C}}^{+}$ + $T{_{\rm C}}^{-}$)/2.

  \begin{table*}[t]
\caption{\label{tab:table1}%
Critical exponents $\beta$, $\gamma$, ${\gamma}'$, and $\delta$ of UTeS and USeS, and those in the mean field theory and various theoretical models with short-range (SR) exchange interactions\cite{privman,fisher0,guillou}. The exponents in UGe$_2$\cite{tateiwa1}, URhGe\cite{tateiwa1}, URhSi\cite{tateiwa2}, UIr\cite{knafo}, and U(Co$_{0.98}$Os$_{0.02}$)Al\cite{maeda} are also shown. Abbreviations: RG-${\phi}^4$, renormalization group ${\phi}^4$ field theory; MAP, Modified Arrott plot; CI, Critical isotherm; KF, Kouvel-Fisher method.}
\begin{ruledtabular}
\begin{tabular}{ccccccccc}
\textrm{}&
\textrm{Method}&
\textrm{$T{_{\rm C}}$(K)}&
\textrm{$\beta$ }&
\textrm{${\gamma}'$($T<T{_{\rm C}}$)}&
\textrm{${\gamma}$($T{_{\rm C}}<T$)}&
\textrm{$\delta$}&
\textrm{Reference}&\\
\colrule
Mean field &&&0.5&\multicolumn{2}{c}{1.0}&3.0&&\\ 
$d$ =  2, $n$ =1 &Onsager solution&&0.125&\multicolumn{2}{c}{1.75}&15.0&\cite{privman,fisher0}&\\ 
$d$ =  3, $n$ =1 &RG-${\phi}^4$&&0.325&\multicolumn{2}{c}{1.241}&4.82&\cite{privman,guillou}&\\ 
$d$ =  3, $n$ =2 &RG-${\phi}^4$&&0.346&\multicolumn{2}{c}{1.316}&4.81&\cite{privman,guillou}&\\ 
$d$ =  3, $n$ =3 &RG-${\phi}^4$&&0.365&\multicolumn{2}{c}{1.386}&4.80&\cite{privman,guillou}&\\ 
\colrule
UTeS&MAP, CI  &84.88 $\pm$ 0.05  &0.309 $\pm$ 0.003 &  \multicolumn{2}{c}{0.998 $\pm$ 0.003} &4.21 $\pm$ 0.04 &this work&\\ 
&KF &85.09  $\pm$ 0.04  &0.315 $\pm$ 0.003 &  \multicolumn{2}{c}{0.996 $\pm$ 0.003}  &&&\\
&Scaling &85.09  $\pm$ 0.03  &0.318 $\pm$ 0.002& 1.03 $\pm$ 0.02 & 1.04 $\pm$ 0.02 &&& \\
\colrule
USeS&MAP, CI & 23.09 $\pm$ 0.03  & 0.300 $\pm$ 0.003 &  \multicolumn{2}{c}{1.00 $\pm$ 0.02 } &4.34 $\pm$ 0.04&this work&\\ 
&KF & 23.18 $\pm$ 0.03 & 0.293 $\pm$ 0.003 &  \multicolumn{2}{c}{0.989 $\pm$ 0.003 }  &&&\\
&Scaling &  23.18 $\pm$ 0.02 & 0.300 $\pm$ 0.002& 1.00 $\pm$ 0.02 & 1.02 $\pm$ 0.02   &&& \\
\colrule
UGe$_2$&Scaling, CI&52.79  $\pm$ 0.02  &0.329 $\pm$ 0.002& 1.00 $\pm$ 0.02 & 1.02 $\pm$ 0.02 &4.16 $\pm$ 0.02&\cite{tateiwa1}& \\
URhGe&Scaling, CI &9.47  $\pm$ 0.01  &0.302 $\pm$ 0.001& 1.00 $\pm$ 0.01 & 1.02 $\pm$ 0.01 &4.41 $\pm$ 0.02&\cite{tateiwa1}& \\
URhSi &Scaling, CI&10.12 $\pm$ 0.02  &0.300 $\pm$ 0.002& 1.00 $\pm$ 0.02 & 1.03 $\pm$ 0.02 &4.38 $\pm$ 0.04&\cite{tateiwa2}& \\
UIr&MAP, CI &45.15 $\pm$ 0.2  &0.355 $\pm$ 0.05 &  \multicolumn{2}{c}{1.07 $\pm$ 0.1} &4.04 $\pm$ 0.05&\cite{knafo}&\\ 
U(Co$_{0.98}$Os$_{0.02}$)Al&MAP, CI &25  &0.33 &  \multicolumn{2}{c}{1.0} &4.18&\cite{maeda}&\\ 
\end{tabular}
\end{ruledtabular}
 \end{table*}

It may be possible to speculate that the 3D Ising universality class below $T_{\rm C}$ is changed to the mean field one above $T_{\rm C}$ in UTeS and USeS. Scaling theory enables us to determine separately the values of ${\gamma}$' for $T<T{_{\rm C}}$ and $\gamma$ for $T{_{\rm C}}<T$. A reduced equation of state close to $T_{\rm C}$ was predicted in the scaling theory as follows\cite{privman}:
 \begin{eqnarray}
 m = f{_{\pm}}{(h)} 
 \end{eqnarray}
Here, $f_{+}$ and $f_{-}$ are regular analytical functions for $T{_{\rm C}} < T$ and $T < T{_{\rm C}}$, respectively. The renormalized magnetization $m$ is defined as $m$ $\equiv$ ${|t|^{-{\beta}}}{M({{\mu}_0}H, t)}$ and the renormalized field $h$ as $h$ $\equiv$ ${{{\mu}_0}}{H}{|t|^{-({\beta}+{\gamma})}}$. Two universal curves are formed in the plot of $m$ vs. $h$ when the correct values of $\beta$, ${\gamma}$', $\gamma$, and $T_{\rm C}$ are chosen. The data in the temperature ranges $t{\,}={\,}|({T}-{T_{\rm C}})/{T_{\rm C}}|{\,}< $ 0.05 for UTeS and 0.08 for USeS are shown in Figs. 5(a) and 5(b), respectively. The analyses yield the values of $T_{\rm C}$ and the critical exponents as $T_{\rm C}$ = 85.09 $\pm$ 0.03 K, $\beta$ = 0.318 $\pm$ 0.002, ${\gamma}'$ = 1.03 $\pm$ 0.02, and $\gamma$ = 1.04 $\pm$ 0.02 in UTeS, and $T_{\rm C}$ = 23.18 $\pm$ 0.02 K, $\beta$ = 0.300 $\pm$ 0.002, ${\gamma}'$ = 1.00 $\pm$ 0.02, and $\gamma$ = 1.02 $\pm$ 0.02 in USeS. This result suggests that the sets of the critical exponents in the two compounds are common below and above $T_{\rm C}$. Note that the magnetic isotherms in the forms of the mean field theory and the 3D Ising model with SR exchange interactions do not form straight lines below and above $T_{\rm C}$ as shown in Figs. 2(a)-2(d). The strongly asymmetric critical region or the change of the universality class across $T_{\rm C}$ can be ruled out.

\section{Discussion}
 Table I shows the critical exponents $\beta$, ${\gamma}'$, $\gamma$, and $\delta$ in UTeS and USeS, and those in mean field theory and various theoretical models with SR exchange interactions of a form $J(r){\,}{\sim}{\,}e^{-r/b}$\cite{privman,fisher0,guillou}. The exponents in the uranium ferromagnetic superconductors UGe$_2$ and URhGe\cite{tateiwa1}, and some uranium ferromagnets URhSi\cite{tateiwa2}, UIr\cite{knafo,sakarya2}, and U(Co$_{0.98}$Os$_{0.02}$)Al\cite{maeda,andreev} are also shown. The sets of the exponents in UTeS and USeS are similar to those of the uranium ferromagnets. The ferromagnetic states of these ferromagnets have strong uniaxial anisotropy. However, the critical exponents of the compounds differ from those in the 3D Ising model with SR exchange interactions. The $\beta$ values are relatively close to those of the 3D models. Meanwhile, the values of $\gamma$ are close to unity, expected one in the mean field theory.

 We discuss the mean-field behavior of the magnetization in the uranium ferromagnetic superconductor UCoGe\cite{huy}. The extent of the asymptotic critical region ${\Delta}{T_{\rm G}}$ where the mean field theory fails can be estimated by the Ginzburg criterion\cite{ginzburg2}. ${\Delta}{T_{\rm G}}$ in three dimensions is given as ${\Delta}{T_{\rm G}}/{T_{\rm C}} = {k_{\rm B}^2}/[32{{\pi}^2}{({\Delta}C)^2}{{\xi_0}^6}]$\cite{chaikin,yelland}. Here, ${\Delta}C$ is the jump of the specific heat at $T_{\rm C}$ in units of erg$\cdot$cm$^{-3}$ K$^{-1}$ and ${\xi}_0$ is the bare correlation length. The value of ${\Delta}{T_{\rm G}}$ for UCoGe was estimated as less than 1 mK using reported ${\Delta}C$ and ${\xi}_0$ values\cite{tateiwa1,huy,stock}. It is natural that the mean field behavior of the magnetization is observed because most of magnetic data might be taken outside the very narrow region around $T_{\rm C}$. The longer magnetic correlation length may be originated from the strong itinerant character of the $5f$ electrons in UCoGe\cite{tateiwa2}. We have previously reported the critical exponents in UGe$_2$\cite{tateiwa1}. The value of ${\Delta}{T_{\rm G}}$ was estimated as $\sim$ 100 K. It can be concluded that the data used for the determination of the critical exponents were taken inside the asymptotic critical region in UGe$_2$. Meanwhile, it is impossible to estimate ${\Delta}{T_{\rm G}}$ for UTeS and USeS since the magnetic correlation length $\xi$ has not been reported so far. In this study, the data for 81.0 K $\le$ $T$ $\le$ 89.0 K and 1.2 T $\le$ ${{\mu}_0}H$ $\le$ 7.0 T in UTeS, and those for 21.4 K $\le$ $T$ $\le$ 24.6 K and 0.4 T $\le$ ${{\mu}_0}H$ $\le$ 3.0 T in USeS are analyzed to determine the critical exponents. If the data up to 7.0 T in USeS are analyzed, they do not form straight lines in the MAP and nor do universal curves in the scaling analysis for any values of the critical exponents. We repeated the analysis and found that the upper limit of the critical region is about 3.0 T for USeS. The consistency in the obtained exponents determined by different methods suggests the reliability of them. We conclude that the data used for the analyses were collected inside the asymptotic critical regions of each compound. 
  \begin{figure}[t]
\includegraphics[width=8.5 cm]{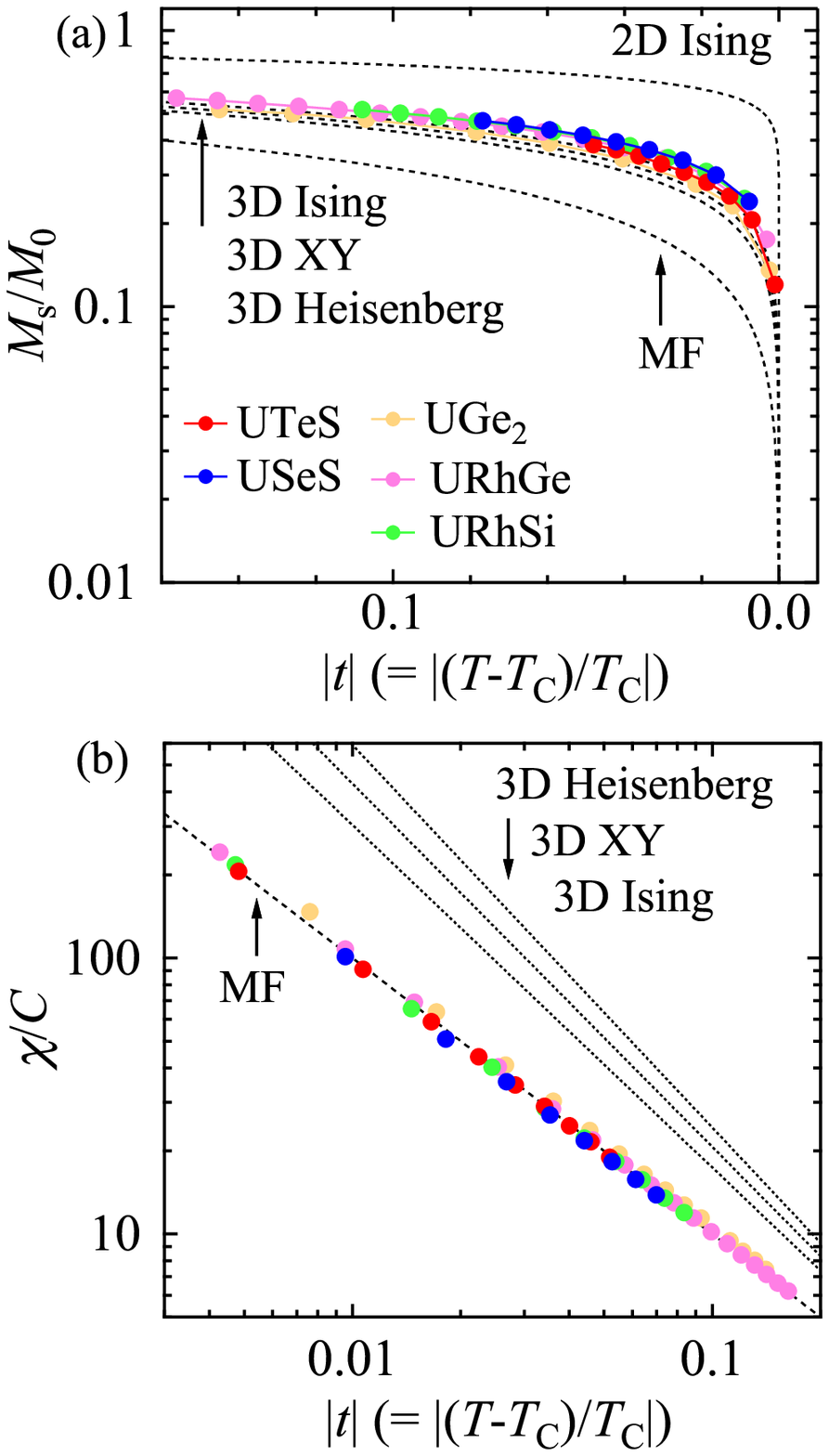}
\caption{\label{fig:epsart}(a) Normalized spontaneous magnetic moment ${M_{\rm s}}/{M_0}$ and (b) magnetic susceptibility ${\chi}/C$ as a function of reduced temperature $|t|$ ($=|(T-{T_{\rm C}})/{T_{\rm C}}|$) for UTeS, USeS, UGe$_2$\cite{tateiwa1}, URhGe\cite{tateiwa1}, and URhSi\cite{tateiwa2}. Dotted lines indicate calculated curves for various theoretical models.  }
\end{figure} 

 Figures 6(a) and 6(b) show the normalized spontaneous magnetic moment ${M_{\rm s}}/{M_0}$ and the magnetic susceptibility ${\chi}/C$ as a function of the reduced temperature $|t|$ ($=|(T-{T_{\rm C}})/{T_{\rm C}}|$) determined from the MAPs in UTeS, USeS, UGe$_2$\cite{tateiwa1}, URhGe\cite{tateiwa1}, and URhSi\cite{tateiwa2}. Here, constants ${M_0}$ and $C$ are obtained from fits to the $|t|$-dependencies of ${M_{\rm s}}$ and $\chi$ with formulas $M_{\rm s}(t)$ = ${M_0}{|t|^{\beta}}$ and ${\chi}(t)$ = $C/{|t|^{\gamma}}$, respectively. The data for the latter three compounds are from our previous studies\cite{tateiwa1,tateiwa2}.  Although the spontaneous magnetization ${M_{\rm s}}$ shows the critical behavior expected for the 3D magnets, the magnetic susceptibility $\chi$ does the mean-field-like behavior. Many theoretical studies have been done for critical phenomena of ferromagnetic transitions in ferromagnetic materials. However, this unusual behavior cannot be explained with previous theoretical approaches. We discuss this issue  from following six viewpoints.

(1) It is well known that the long-range nature of magnetic exchange interactions affects the critical behavior of the magnetization around $T_{\rm C}$. The theoretical values of the critical exponents in Table I are those of theoretical models with short-range (SR) exchange interactions of a form $J(r){\,}{\sim}{\,}e^{-r/b}$\cite{privman,fisher0,guillou}. The strength of the exchange interaction $J(r)$ decreases rapidly with increasing distance $r$. When the range of the exchange interaction becomes longer, the critical exponents of the models shift toward those of the mean field theory. This problem was studied by Fischer {\it et al.} with a renormalization group approach for systems with the exchange interaction of a form $J(r)$ $\sim$ $1/r^{d+{\sigma}}$\cite{fisher1}. Here, $\sigma$ is the range of exchange interaction and $d$ is the dimension of the system. They showed that the model is valid for ${\sigma}{\,}<{\,}2$ and derived a theoretical formula for the exponent $\gamma$ = $\Gamma{\{}{\sigma}, d, n{\}}$. Here, $n$ is the dimension of the order parameter and the function $\Gamma$ is given in Ref. 34. We calculate the critical exponents using the formula and scaling relations for different sets of ${\{}d: n{\}}$ ($d$, $n$ = 1, 2, 3) in order to reproduce the exponents in the uranium ferromagnets. However, there is no reasonable solution of $\sigma$.
  
(2) The critical phenomenon of the magnetization is also affected by the classical dipole-dipole interaction as has been studied for rare earth metal gadolinium ($T_{\rm C}$ = 292.7 K and the spontaneous magnetic moment $p_{\rm s}$ = 7.12 ${{\mu}_{\rm B}}$/Gd)\cite{srinath}. The interaction may not have a strong effect on critical phenomena in uranium ferromagnets since the strength of the effect is proportional to $p_{\rm s}^2$\cite{fisher2}. The value of $p_{\rm s}$ is determined as 1.62 ${{\mu}_{\rm B}}$/U at 5.0 K for UTeS and 1.09 ${{\mu}_{\rm B}}$/U at 2.0 K for USeS. These are much smaller than that of Gd. Moreover, the exponents in the uranium ferromagnets are not consistent with those of theoretical studies for the critical phenomenon associated with the isotropic or anisotropic dipole-dipole interaction\cite{frey1,frey2}. 
  
(3) We discuss the critical exponents from the viewpoint of the local moment magnetism. The ferromagnetic states in the uranium ferromagnets can be regarded as the anisotropic 3D Ising system or the anisotropic next-nearest-neighhor 3D Ising (ANNNI) system. However, the critical exponents in the uranium ferromagnets are not consistent with those obtained in numerical calculations for the two systems\cite{yurishchev,murtazev}.
  
(4) The temperature dependencies of the spontaneous magnetic moment $p_{\rm s}$ and the magnetic susceptibility $\chi$ obtained analytically or numerically in the spin fluctuation theories are not consistent with those in the uranium ferromagnets\cite{moriya,takahashi}. The spin fluctuation theories cannot be applied to physical phenomena in the asymptotic critical region.

 (5) The unconventional critical phenomenon in UGe$_2$ and URhGe has been discussed by Singh, Dutta, and Nandy with a nonlocal Ginzburg-Landau model focusing on magnetoelastic interactions\cite{singh1}. It was claimed that their calculated results are comparable with those of UGe$_2$ and URhGe. It is hoped that the almost mean-field behavior of $\chi$ is completely reproduced. 

 The itinerant picture of the $5f$ electrons is basically appropriate to describe the ferromagnetism in uranium intermetallic compounds\cite{tateiwa3}. Meanwhile, the dual nature of the $5f$ electrons in UGe$_2$ has been experimentally suggested in the Muon spin rotation spectroscopy\cite{yaouanc,sakarya}. The concept of the duality of the $5f$ electrons has been a basis in theoretical studies for the superconductivity in UGe$_2$\cite{troc2}, URhGe\cite{hattori2}, and UPd$_2$Al$_3$\cite{thalmeier}. Previously, we pointed out relevance between the dual nature of the $5f$ electrons and the novel critical behavior of the magnetization in UGe$_2$, URhGe, and URhSi\cite{tateiwa1,tateiwa2}. A novel critical phenomenon can be expected due to two correlation lengths of the localized and itinerant components of the $5f$ electrons and a Hund-type coupling between them. However, this scenario cannot be applied to UTeS and USeS with the localized $5f$ electrons. The soft X-ray photoelectron spectroscopy showed that the $5f$ level is situated about 750 meV below the Fermi energy in UTeS\cite{takeda}. The present study shows that the novel critical phenomenon of the ferromagnetic transition is observed not only in the uranium intermetallic compounds where the ferromagnetism is carried by the itinerant $5f$ electrons but also in UTeS and USeS with the localized $5f$ electrons. The large negative magnetoresistance in UTeS and USeS, and the uranium ferromagnetic superconductivity in UGe$_2$ and URhGe are observed in the vicinity of the ferromagnetism characterized by the novel critical exponents.

 It has been long thought that the $p$-wave superconductivity in the uranium ferromagnetic superconductors is driven by longitudinal spin fluctuations developed in the vicinity of the ferromagnetic state described with the 3D Ising model. Meanwhile, our studies suggest that the ferromagnetic correlation between the $5f$ electrons differs from that of the 3D Ising system in the uranium ferromagnets including the superconductors. Recent uniaxial experiment on URhGe and its theoretical interpretation suggest that a pairing mechanism other than that driven by Ising-type longitudinal fluctuations take a certain role for the superconductivity\cite{braithwaite,mineev}. It would be interesting to study the dynamical magnetic property of the uranium dichalcogenides. It was claimed that the superconductor UTe$_2$ is on the verge of the ferromagnetism since the critical exponents are close to values expected for a ferromagnetic quantum critical point\cite{ran1,kirkpatrick}. Magnetic fluctuations observed in muon spin relaxation/rotation ($\mu$SR) measurements on UTe$_2$ may take an important role for anomalous behaviors of the upper critical field $H_{\rm {c2}}$ or the unconventional superconducting order parameter with point nodes suggested from the thermal transport, heat capacity and magnetic penetration depth measurements\cite{sundar,ran2,metz}. The group of the uranium dichalcogenides would be an interesting platform for the study of both the large magnetoresistance and the superconductivity in terms of the ferromagnetic correlation between the $5f$ electrons.

\section{Summary}
 In summary, we study the novel critical behavior of magnetization in uranium semimetal UTeS and semiconductor USeS exhibiting a large transverse magnetoresistance around the ferromagnetic transition temperatures. The critical exponents in the two compounds differ from those in the 3D Ising model with short-range exchange interactions in spite of uniaxial magnetic anisotropy in the ferromagnetic states. The critical exponents are similar to those in uranium ferromagnetic superconductors UGe$_2$ and URhGe, and some uranium ferromagnets URhSi, UIr and U(Co$_{0.98}$Os$_{0.02}$)Al. The universality class for the ferromagnetic transition in UTeS and USeS may belong to the same one for the uranium ferromagnets. The novel critical phenomenon of the ferromagnetic transition appears not only in the uranium intermetallic ferromagnets with the itinerant $5f$ electrons but also in UTeS and USeS with the localized $5f$ electrons. There is similarity between the large magnetoresistance in UTeS and USeS, and the superconductivity in UGe$_2$ and URhGe of their closeness to the ferromagnetism characterized by the novel critical exponents.

 \section{Acknowledgments}
 This work was supported by Japan Society for the Promotion of Science (JSPS) KAKENHI Grant No. JP16K05463, JP16KK0106, and JP17K05522.

\bibliography{apssamp}

\end{document}